\documentclass[preprint,12pt]{elsarticle}

\usepackage{amssymb}

\usepackage{chemformula}
\usepackage[referable]{threeparttablex}
\usepackage{booktabs}
\usepackage{lineno}
\newcommand{\specialcell}[2][c]{%
  \begin{tabular}[#1]{@{}c@{}}#2\end{tabular}}

\journal{Comput. Struct. Biotechnol. J.}

\begin{document}

\begin{frontmatter}


\title{Poly-Sarcosine and Poly(ethylene-glycol) interactions with proteins investigated using molecular dynamics simulations}
 \author[phys]{Giovanni Settanni\corref{cor1}}
 \ead{settanni@uni-mainz.de}
 \ead[url]{http://www.staff.uni-mainz.de/settanni/}
 \cortext[cor1]{Corresponding author}
 \author[phys]{Timo Sch\"{a}fer} 
 \author[chem]{Christian Muhl}
 \author[chem]{Matthias Barz}
 \author[phys]{Friederike Schmid}
 \address[phys]{Institut f\"{u}r Physik, Johannes Gutenberg University, Mainz, Germany}
 \address[chem]{Institut f\"{u}r Organische Chemie , Johannes Gutenberg University, Mainz, Germany}
\begin{abstract}
Nanoparticles coated with hydrophilic polymers often show a reduction in unspecific interactions with the biological environment, which improves their biocompatibility. The molecular determinants of this reduction are not very well understood yet, and their knowledge may help improving nanoparticle design.
Here we address, using molecular dynamics simulations, the interactions of human serum albumin, the most abundant serum protein, with two promising hydrophilic polymers used for the coating of therapeutic nanoparticles, poly(ethylene-glycol) and poly-sarcosine. By simulating the protein immersed in a polymer-water mixture, we show that the two polymers have a very similar affinity for the protein surface, both in terms of the amount of polymer adsorbed and also in terms of the type of amino acids mainly involved in the interactions. We further analyze the kinetics of adsorption and how it affects the polymer conformations. Minor differences between the polymers are observed in the thickness of the adsorption layer, that are related to the different degree of flexibility of the two molecules. In comparison poly-alanine, an isomer of poly-sarcosine known to self-aggregate and induce protein aggregation, shows a significantly larger affinity for the protein surface than PEG and PSar, which we show to be related not to a different patterns of interactions with the protein surface, but to the different way the polymer interacts with water.
\end{abstract}
\begin{keyword}
PEG \sep poly-peptoid \sep poly-sarcosine \sep MD simulation \sep nanoparticle protein corona
\end{keyword}
\end{frontmatter}
\section{Introduction}
\label{intro}
Huge steps in the production of nanosized materials with specific functionalities have opened the way to a vast variety of applications, one of which is their use as drug delivery systems \cite{cho2008}. A wide spectrum of nanoparticles are being tested for their capabilities to load different kinds of cargos, remain soluble in the biological milieu (e.g. blood, mucosa, etc.), evade secretion and immune response, target specific tissues, and specifically release their cargo. The efficiency of most of these processes depends on how the nanoparticle surface interacts with the biological medium in which it is introduced. These interactions determine the composition of the layer of biological material (protein corona) that forms around nanoparticles as they come in contact with an organism. It is precisely the protein corona that has been shown to determine the fate of the nanoparticle in the host organism, in terms for example of circulation time, cell uptake or immunogenicity\cite{monopoli2012,lesniak2012}.

Coating of the nanoparticle surface with specific materials represents a very common way to control the protein corona composition. A large variety of coating strategies have been tested in the last few years. A particular strategy consist in the exploitation of the so called "stealth" effect, that is the capacity of  certain materials, especially polymers, to reduce unspecific interactions with the surrounding biological milieu. Nanoparticles coated with these polymers show reduced protein corona formation and in some cases reduced toxicity. Poly(ethylene-glycol) (PEG) is the most common of the polymers showing a stealth effect, but recently other polymers have been investigated like Poly-phosphonates\cite{Schoettler2016}, poly(N-(2-hydroxypropyl) methacrylamide)\cite{Nuhn2014} or polypeptoids like poly-sarcosine (PSar)~\cite{Weber2016}.

The molecular determinants of the stealth effect are not yet very well understood. One qualitative explanation is that the highly hydrophilic nature of the stealth polymers helps and creates a layer of water around the coated material which, then, reduces the interactions with the surrounding environment. In reality, however, protein coronas form also in the presence of stealth coatings\cite{Schoettler2016}. So the stealth effect may not only be related to a generic reduction in the amount of adsorbed proteins but also in the kind of interactions that the coating makes with those proteins.

A direct experimental characterization of those molecular interactions is made difficult by the small length scales and the complexity of the systems, which is often not easily and completely controllable. An {\it in silico} approach, on the other hand, allows for a reduction of the complexity of the problem by focusing only on selected aspects and could, in principle, provide a high resolution picture of the involved mechanisms. Coarse-grained models of the adsorption of biomolecules on nanoparticle have been adopted to study the dynamics~\cite{Vilaseca2013} and the structural outcomes of the process~\cite{Heller2017} and are definitively necessary for a streamlined approach to a fast and accurate evaluation of nanomaterial toxicity~\cite{Lopez2017}. These models need a solid and accurate basis to build upon, which is represented by atomistic models.
During the last 40 years increased availability of computer resources and improvements in the accuracy of the force fields have made possible the atomistic molecular dynamics (MD) simulation of biological systems (proteins, nucleic acids, lipids sugars, etc.) of the size of tens to hundreds nanometers on time scales of hundreds to thousand nanoseconds (even milliseconds on specially designed architectures)\cite{fastfolding}.
Further recent improvements in force field development, namely the availability of force field parameters describing materials including silicates, metals, oxides as well as graphitic materials~\cite{INTERFACE,Dharmawardhana2017,Wright2013} or polymers~\cite{Lee2008,sarcosineff, Prakash2018,cgenff}, and compatible with those available for biological matter, like proteins, lipids, nucleic acids and sugars, have made possible the use of MD to study the interface between biological molecules and material surfaces~\cite{kohler2015a, Sola-Rabada2018,Hildebrand2018,Hughes2017} as reviewed in ref.~\cite{ozboyaci_modeling_2016} and~\cite{heinz_simulations_2016}. 

In the case of PEG, for example, our MD simulations revealed that, irrespective of the protein being considered, the PEG density around each amino acid depends mainly on its type, with negatively charged residue showing the lowest densities and non-polar residues showing the highest. In other words effective attractive interactions between PEG and non-polar residues were observed in the simulations, while the effective interactions between PEG and negatively charged residues were mainly repulsive~\cite{settanni2017}. 
From those observations we have derived a model that describes the interactions of proteins with densely PEGylated nanoparticles using only the amino acid composition of the protein surface. We then applied the model to a large set of blood proteins, for which the three-dimensional structure has been determined, and verified a good correlation between the expected PEG density around the protein as derived from the model and the adsorption free energy of the proteins on PEGylated nanoparticles measured using mass spectroscopy experiments~\cite{settanni2017}. After noting that the adsorption free energies of the blood proteins on the PEGylated nanoparticles are highly correlated to those measured on nanoparticles densely grafted with poly-phosphonates, we showed that exactly the same model obtained for protein interactions with the PEGylated nanoparticle can be applied to the poly-phosphonated ones~\cite{settanni2017}. Further analysis of the simulations also allowed to measure the differential binding coefficients of several proteins in PEG water mixtures\cite{Settanni2017a}. 

\begin{figure}
\caption{\label{poly_struc} Chemical formulas of PEG (left), PSar (middle) and PAla (right).  }
\begin{center}
\includegraphics[width=8.5cm]{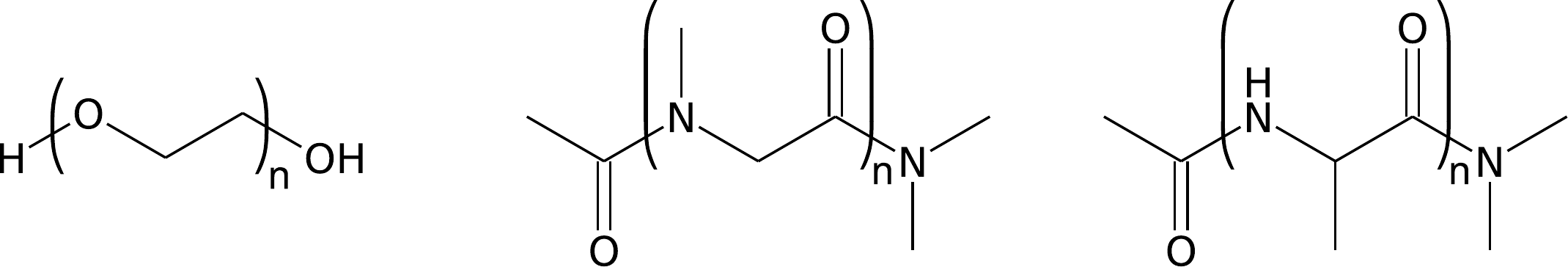}
\end{center}
\end{figure}

Following the results obtained for PEG and poly-phosphonate, here, we have extended our analysis to PSar. PSar is a poly-peptoid (the monomer is similar to alanine but with the methyl residue bound to the backbone nitrogen atom rather than to the C$_\alpha$ atom, see Fig.~\ref{poly_struc}), which, like PEG, can help reduce unspecific interactions with proteins~\cite{Birke2018,Weber2018,Klein2018,Yoo2018,Ostuni2001}, but, unlike PEG, it can be metabolized by the organism\cite{polypeptoids}. These facts make PSar a very promising substitute for PEG, which, on the other hand, in some cases has been shown to induce immune reactions\cite{Ishida2006}. Notwithstanding the  potential applications, the way PSar interacts with biological macromolecules and proteins, in particular,   is not yet well understood and, to our knowledge, has not been addressed at the molecular level, especially with theoretical tools. Here we have confronted this issue  by simulating  a representative blood protein, Human Serum Albumin(HSA) (in the case of PEG the pattern of interactions are basically independent of the protein being considered~\cite{settanni2017}), immersed in a PSar/water mixture at physiological pH and ionic strength. We have then compared the observed behavior with the one of PEG under similar conditions, as well as poly-alanine(PAla), a polymer with the same chemical composition as PSar but a remarkably different behavior, due to its tendency to aggregate\cite{Giri2007}. In addition to what done in ref.~\cite{settanni2017} for PEG only, here we have carefully characterized for all the three polymers the dynamics of the adsorption process on the protein surface and the effect of adsorption on the polymer conformations.

\section{Methods}
Molecular dynamics simulations were carried out using the program NAMD\cite{NAMD} with the CHARMM force field\cite{CMAP}. A base time step of 1fs was used. Direct space non-bonded interactions were cut-off at 1.2nm with a switch function smoothing them out from 1.0 to 1.2nm. The neighbor list cutoff was 1.4nm.  A specially designed cell-list algorithm was used to speed up neighbor search\cite{allentildesley}. Long range electrostatic interactions were treated using the smooth particle mesh Ewald (PME) method\cite{Essmann1995}. A multiple time step scheme was used\cite{multipletimestep}, so that non-bonded interactions were computed every 2 steps and long range electrostatics every 4. Water molecules were simulated explicitly using the TIP3P model\cite{TIP3P}. A Langevin piston\cite{Martyna1994,Feller1995} was used to control the temperature and the pressure of the system at 300K and 1atm, respectively. 
Similar to the protocol used with PEG~\cite{settanni2017} the simulation setup for PSar-protein and PAla-protein simulations, consisted in an initial preparation and equilibration of the polymer/water mixture at different concentrations of the polymer, followed by insertion of the protein and removal of all the molecules (water or polymer) within 0.21~nm from the protein. The protein used here was HSA (pdbid 1AO6 \cite{1AO6}) which is the most abundant protein in the blood, and consists of 578 residues (179 charged, 168 polar uncharged and 231 hydrophobic) with a total charge of -15e. All the  systems were later neutralized and set at physiological ion concentration \ch{[NaCl]=150mM} by replacing few water molecules with sodium and chlorine ions. This led to cubic simulation boxes as reported in tab.~\ref{simdata}. The prepared systems were  minimized for 10000 steps using steepest descent with positional restraints on the heavy atoms of the protein and equilibrated in NPT for 1~ns gradually releasing the constraints and for another 1~ns without constraints. In addition to the CHARMM force field~\cite{CHARMM}, a specific force field for PSar was adopted~\cite{sarcosineff}. Short polymer chains of 4 monomers were used to enhance diffusivity of the polymer over the protein surface. Autocorrelation times of the protein polymer interactions were estimated at around 10ns \cite{settanni2017}.   For the production phase, four or five independent runs were carried out each for 200ns. The first 10ns of each trajectory were not used for the analysis to allow proper diffusion of the polymer around the protein.  The trajectories were  analyzed using VMD~\cite{VMD} and WORDOM~\cite{Seeber2007}.
A list of the simulation data used is provided in Tab.~\ref{simdata}.
\begin{table}
\caption{\label{simdata}List of the analyzed simulations}
\begin{threeparttable}
\begin{tabular}{llcccccccc}
\toprule
System & Box size (\AA) & N. Atoms & \specialcell{Polymer\\length} & \specialcell{Polymer\\molecules} &\specialcell{Concentration\\(g/ml)} & \specialcell{Simulation\\time (ns)}\\ 
\midrule
PEG1\tnote{a}&  98.8 & 100881 & 4 & 214 & 0.08 & 4 x 200 \\ 
PEG2\tnote{a}& 98.2 & 99301 & 4 & 292& 0.11  & 4 x 200 \\ 
PEG3\tnote{a}& 108.6& 134134 & 4 & 424 & 0.12  & 4 x 200 \\ 
PEG4\tnote{a}& 109.0& 134778 & 7 & 88 & 0.04  & 5 x 200 \\ 
PEG5\tnote{a}& 118.2 & 172541 & 4 & 560 & 0.12 & 5 x 100 \\ 
\midrule
PSar1& 98.5 & 99894 & 4 & 83 & 0.060 & 4 x 200 \\ 
PSar2& 98.5 & 100128 & 4 & 103 & 0.074& 5 x 200 \\ 
\midrule
PAla1& 98.3 & 99331 & 4 & 66 & 0.047 & 5 x 200 \\ 
PAla2& 98.3.2 & 99413 & 4 & 76& 0.055 & 5 x 200 \\ 
\bottomrule
\end{tabular}
\begin{tablenotes}
\item[a] Trajectories from ref.~\cite{settanni2017}.
\end{tablenotes}
\end{threeparttable}
\end{table}

\section{Results and Discussion}
\begin{figure}
\caption{\label{show_traj} Snapshots from the simulations of PEG$_7$ (PEG4 A,B,C), Sar$_4$ (PSar2 D, E, F) and Ala$_4$ (PAla2 G, H, I). The snapshots at 10~ns  (A, D, G), 100ns (B, E, H) and 200ns (C, F, I) for one of the runs are reported. The protein is represented as cartoon . The conformations of the polymer atoms within 0.5nm of the protein are reported as black lines. All the polymer conformations sampled in the 1ns following the snapshot are shown superimposed, to provide information about their variability.  }
\begin{center}
\includegraphics[width=13.5cm]{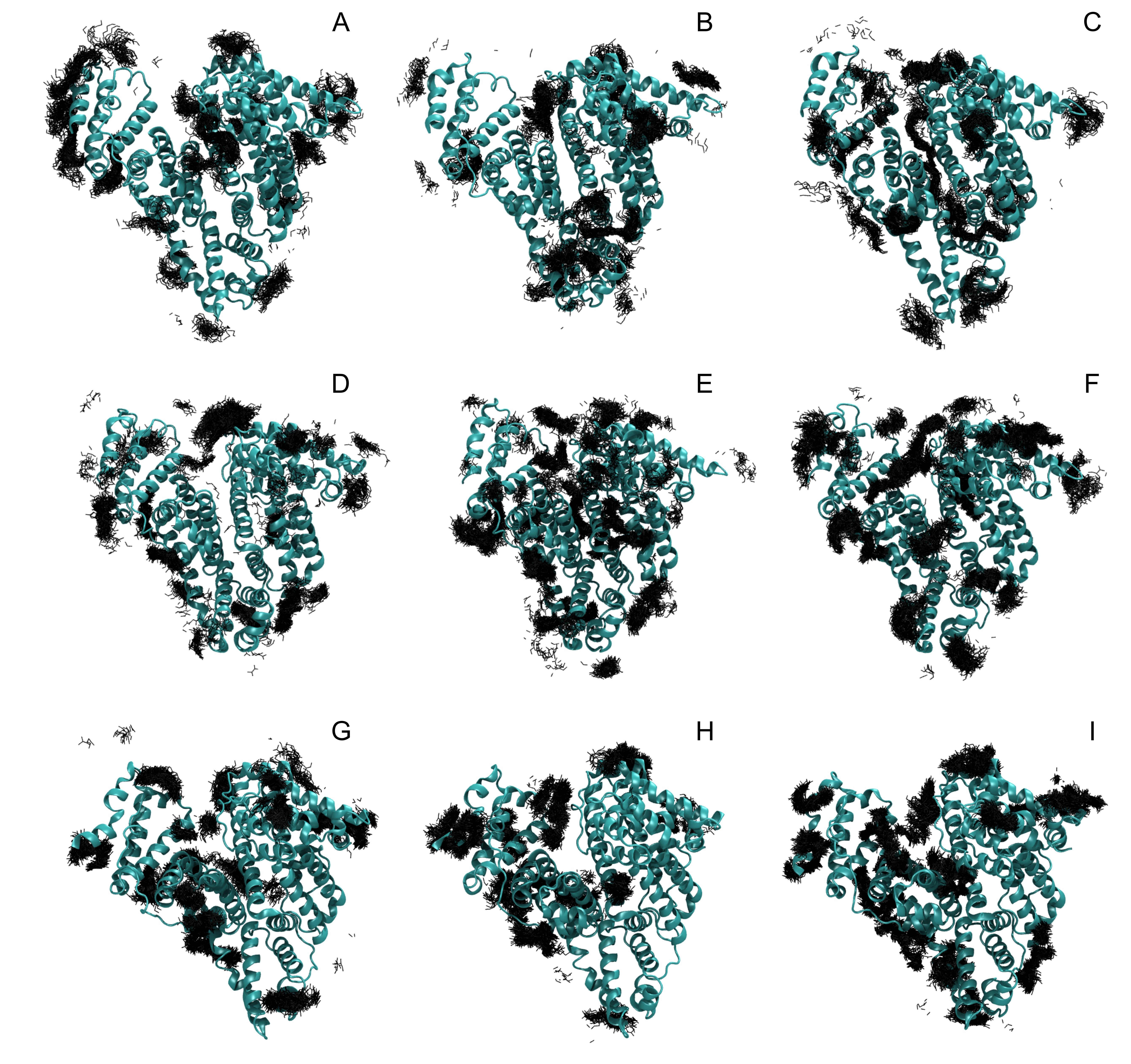}
\end{center}
\end{figure}

The structure of HSA is mostly unaffected by the presence of the polymers in the solvent. The $C_\alpha$-RMSD with respect to the equilibrated crystallographic structure mostly oscillates around or below 0.35~nm in all tested water/polymer mixtures, which is expected for proteins of similar size during the simulated timescales. In the course of the trajectories the polymers diffused around the protein and sampled a variety of configurations (Fig.~\ref{show_traj}).  

\begin{figure}
\caption{\label{hsa_gofr} The radial distribution function of the polymer atoms  around the protein surface in the simulations as indicated in the legend.  }
\begin{center}
\includegraphics[width=8.5cm]{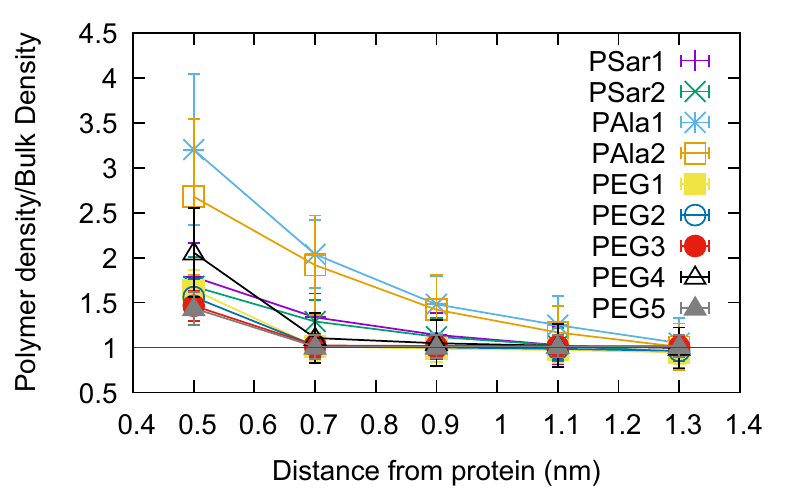}
\end{center}
\end{figure}

The simulations revealed that the density of polymer atoms decreases with the distance from the protein in all the studied cases (Fig.~\ref{hsa_gofr}). In the case of PEG the effect extends only up to 0.5~nm from the protein surface, in the case of PSar it is still detectable at 0.7~nm and in the case of PAla it extends further up to 1.1~nm. These data indicate the presence of attractive interactions between the protein surface and the polymers. In case of PEG$_4$, PEG$_7$ and PSar, whose average radius of gyration in the simulations is 0.34$\pm$0.03~nm, 0.49$\pm$0.06~nm and 0.43$\pm$0.03~nm respectively, the data are compatible with single molecules attaching to the protein surface. In the case of PAla, the effect goes beyond the thickness of a single molecule, whose radius of gyration is 0.41$\pm$0.05~nm, and hints to the attachment of clusters of molecules on the protein surface. 

\begin{figure}
\caption{\label{langmuir}Density of polymer heavy atoms in a 0.5nm-thick layer of solvent around the protein as a function of overall polymer concentration in the simulation box. PEG, PSar and PAla are represented in red, green and blue, respectively. The error bars represent standard deviations from the simulations. The continuous lines represent fits of the Langmuir-like adsorption model eq.~\ref{langmuir_isotherm}.}
\begin{center}
\includegraphics[width=12.cm]{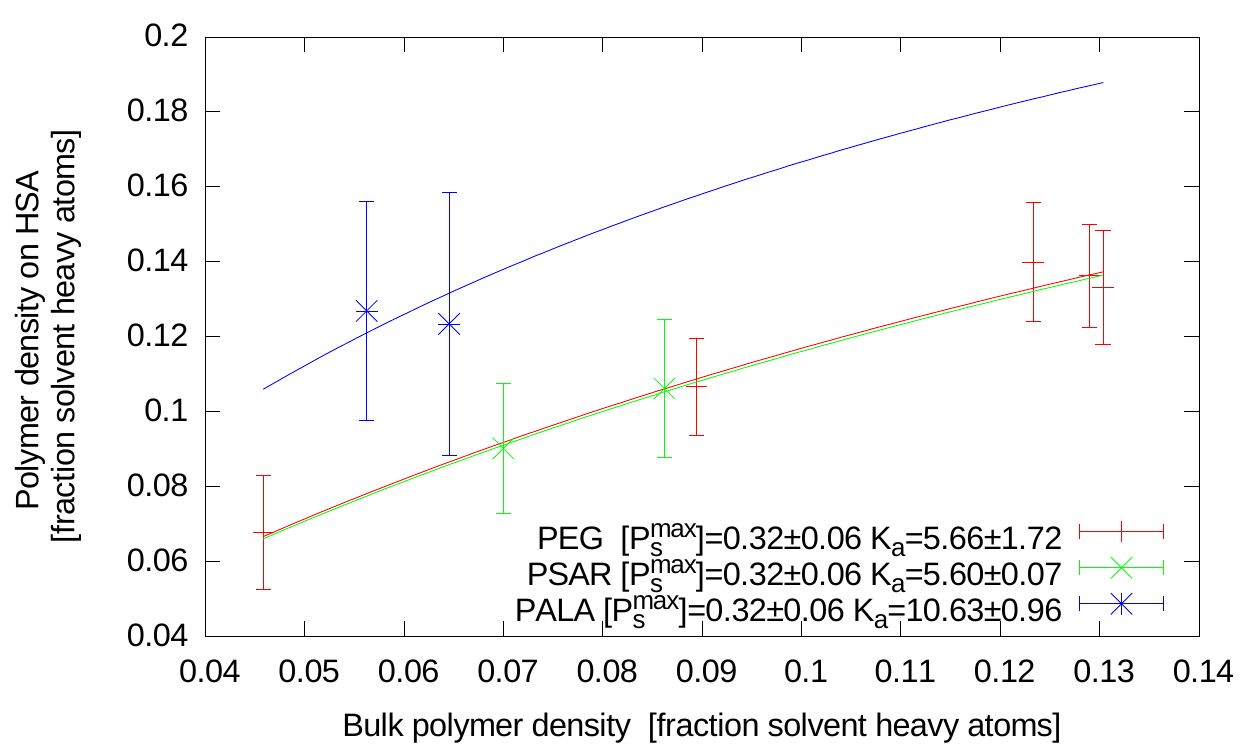}
\end{center}
\end{figure}

The density of the polymer can also be monitored by measuring the fraction of polymer (heavy) atoms in the vicinity of the protein (in a 0.5nm-thick layer around the protein) with respect to the total number of heavy atoms (water and polymer) \cite{settanni2017, Settanni2017a}. This number, normalized to the fraction of polymer heavy atoms in the solvent, can be used to quantify more precisely the affinity of the polymer for the protein surface (Fig.~\ref{langmuir}). To do so we assumed a very simplified Langmuir-like model of adsorption, where a finite number of polymer binding sites is present on the protein surface, each of which can bind a polymer (heavy) atom with approximately the same affinity. We assume, then, that each polymer atom can undergo an adsorption reaction of the form \ch{A_s + P_s <=> A_P}, where \ch{A_s} is a generic (heavy) atom of the polymer in solution, \ch{P_s} is an empty binding site on the protein surface and \ch{A_P} is the polymer atom adsorbed on the protein surface.  This model leads to a Langmuir-like isotherm~\cite{Masel1996}:
\begin{equation}
\label{langmuir_isotherm}
\ch{[A_P]}=\ch{[P_s^{max}]} \frac{\ch{K_a.[A_s]}}{\ch{K_a}+\ch{[A_s]}}
\end{equation}
\noindent where  \ch{[A_s]} is the concentration of polymer atoms in the solvent (i.e., the fraction of polymer heavy atoms), \ch{[A_P]} is the concentration of polymer atoms in the vicinity of the protein (again, measured as the fraction of polymer heavy atoms), \ch{[P_s^{max}]} is the maximum concentration of binding sites on the protein surface, and \ch{K_a} is the equilibrium constant of the adsorption reaction. We can then fit eq.~\ref{langmuir_isotherm} to the simulation data collected for PEG on HSA, which results in  \ch{[P_s^{max}]}$=0.32\pm.05$ and \ch{K_a} $= 5.66\pm1.49$ (fig.~\ref{langmuir}). In the case of PAla and PSar, where the the number of data points available in fig.~\ref{langmuir} is small, we assumed the same \ch{[P_s^{max}]} as for PEG and we fitted the \ch{K_a}. This resulted in \ch{K_a}$=5.00\pm0.05$ and $10.63\pm0.68$ for PSar and PAla, respectively. This clearly shows that PAla has a significantly larger affinity for the HSA surface than PSar and PEG. PSar and PEG have, instead, a similar affinity. The assumption that the maximum density of polymer binding sites is the same for all the polymers is rather crude, especially comparing PAla with the hydrophilic polymers PSar and PEG. By fitting the \ch{[P_s^{max}]} with the same \ch{K_a} as for PEG, PAla data would show a larger concentration of binding sites on the protein surface, thus confirming that the affinity of PAla for the protein surface is larger than the other two polymers.

\begin{figure}
\caption{\label{occupancy}Cartoon of HSA showing the regions with a polymer density higher than a cutoff, taken to be twice as the overall density of polymer in the box. Data for PEG, PSar and PAla are shown in (a), (b) and (c),  respectively. For these data the simulations PEG1, PSar2, PAla2 have been used, where the overall density of polymer in the simulation box is approximately similar. High water density regions, not shown here for clarity, approximately cover the protein surface areas not occupied by the high polymer density regions. Inset: a snapshot of the PAla2 simulation showing the aggregated clumps of PAla (cyan-blue-red) around the protein (yellow).}
\includegraphics[width=\columnwidth]{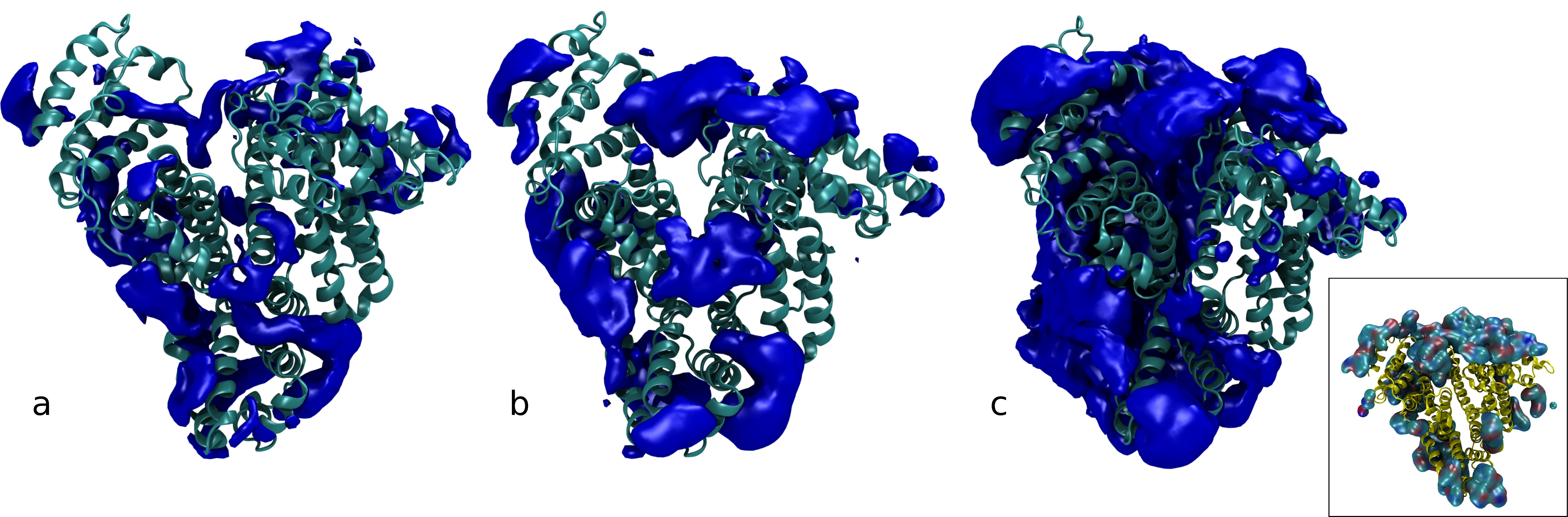}
\end{figure}

We then compared the distribution of the polymers on the protein surface, in order to detect differences in the adsorption patterns. We measured the density of polymer per {nm$^3$} in the simulation box averaged over the whole set of simulations after aligning all the snapshots to the initial conformation of the protein. These data can then be visualized using VMD\cite{VMD} (fig.~\ref{occupancy}) and they show that, while for PSar and PEG the distributions are relatively similar and relatively symmetric with respect to the protein center of mass, in the case of PAla the distribution is significantly skewed  on one side and covers a wider surface region. The behavior of PAla depends on its propensity to aggregate which is expected. Namely, in PAla simulations PAla aggregates tend to form (Fig.~\ref{occupancy} inset). This point will be further discussed below.

\begin{figure}
\caption{\label{fig1} Comparison of the polymer/water ratios measured around each amino acid type (indicated using the single-letter code) for each pair of polymers considered. Although those values show high correlations (r-values), those from PAla are systematically larger than the others, indicating larger affinity for the protein surface. The lines indicate the identity function. The data come from simulations PEG1, PSar2 and PAla2}
\begin{center}
\includegraphics[width=8cm]{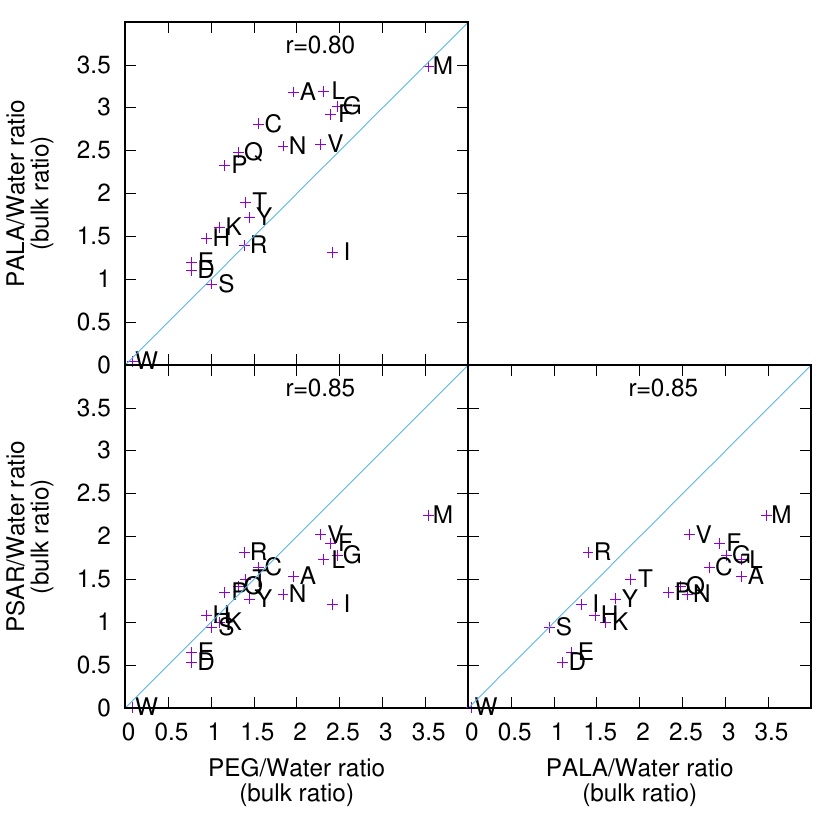}
\end{center}
\end{figure}

In earlier work, we showed that the affinity of PEG and poly-phosphonate for the protein surface can be broken down into the contributions coming from each amino acid on the protein surface\cite{settanni2017}. These contributions can be computed by measuring the ratio between polymer heavy atoms and water oxygen atoms in a 0.5~nm shell around each amino acid type and comparing it to the overall ratio (bulk ratio) in the simulation box. We have measured these quantities for PSar and PAla and compared them to those obtained for PEG.
The results show that the affinity of the various amino acids for PSar are similar to those measured for PEG (Fig.~\ref{fig1}), which agrees with what observed already in data showed so far. The similarity holds particularly well for the polar amino acids at low polymer/water ratios,  while the non-polar amino acids show higher affinity for PEG than PSar. In the case of PAla, the residue specific polymer/water ratios also correlate with those measured for PEG and PSar, however they are systematically larger. This confirms that the affinity of PAla for the protein surface is larger than the other two polymers, and indicates that this is not due to a different interaction pattern with the surface amino acids, but simply due to stronger interactions with the same amino acids. 

\begin{figure}
\caption{\label{atm_p_mol} Distribution of the number of atoms adsorbed on the protein surface for each polymer molecule (polymer heavy atoms within 0.5~nm of protein heavy atoms). PEG$_7$, PSar$_4$ and PAla$_4$ molecules contains 22, 26 and 26 heavy atoms in total, respectively. PEG is mostly found either completely adsorbed or desorbed, PSar and PAla show a variety of partially adsorbed conformations, due to their bulkier and less flexible structure. Only data for the heaviest polymers are shown as they are more easily comparable.}
\begin{center}
\includegraphics[width=8cm]{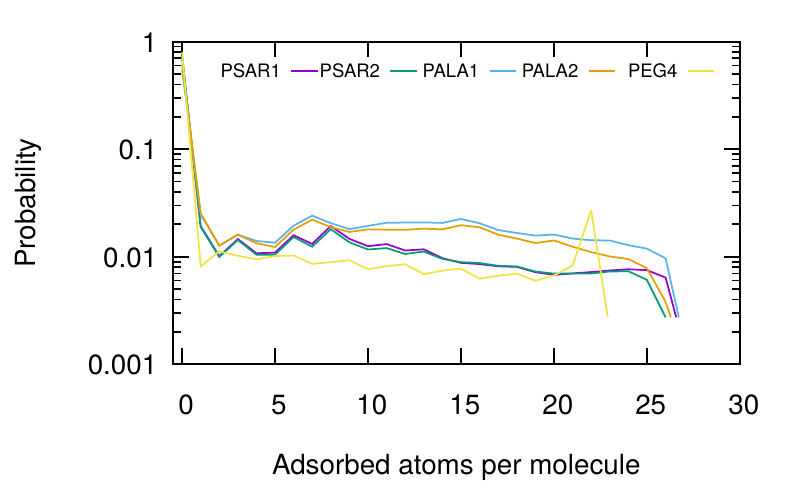}
\end{center}
\end{figure}

\begin{figure}
\caption{\label{shape_comp} Distribution of end to end distance (A), radius of gyration (B) and aspect ratio (C) for the polymers in the various simulations (dashed lines). The solid lines represent the same distributions for adsorbed molecules only. }
\begin{center}
\includegraphics[width=8cm]{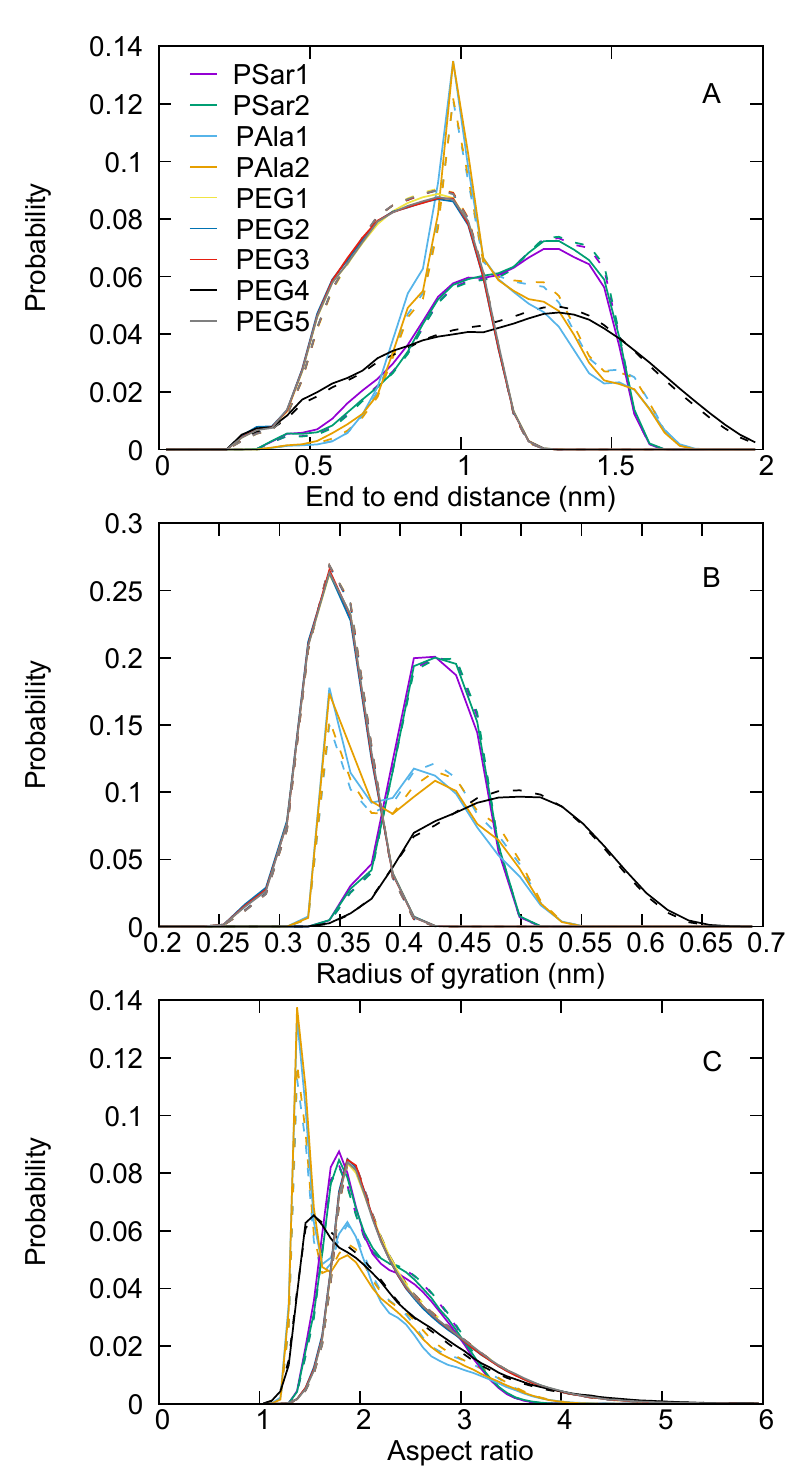}
\end{center}
\end{figure}

After defining as adsorbed the heavy atoms of the polymer within 0.5~nm from the heavy atoms of the protein, we measured the distribution of adsorbed atoms per polymer molecule along the simulations (Fig.~\ref{atm_p_mol}). This shows that in the case of PEG, molecules tend to be either completely adsorbed or completely desorbed. In the case of PSar or PAla, on the other hand, the molecules show a variety of partially adsorbed states. We further investigated the shape of the polymers by measuring their end-to-end distance, radius of gyration and aspect ratio (measured by the square root of the ratio between the largest and smallest eigenvalue of the inertia tensor). These data (Fig.~\ref{shape_comp}) show that PEG is a more flexible molecule than PSar and PAla accessing a broader distribution of conformations. These differences in shape and adsorption behavior can be explained by the different degree of branching of PEG and PSar (or PAla) which imparts PEG a larger flexibility. PEG's very short persistence length of less than 0.4~nm, allows it to adapt to the underlying protein surface. The branched nature of PSar and PAla provides them with additional rigidity and bulkiness, which sometime may prevent all the atoms of the molecules to reach the 0.5~nm region around the protein. The data further show that PSar has on average a more elongated conformation with respect to PAla, with larger aspect ratio, radius of gyration and end to end distance. PAla in particular shows the presence of two main conformational states, as evidenced by the presence of two peaks in the distribution of gyration radii and aspect ratios. One state is more compact and the other one is more elongated. The population of the adsorbed PAla molecules is slightly richer in the compact state than the elongated one (Fig.~\ref{shape_comp}), while in the case of PSar and PEG, no large shape change is observed upon adsorption. 

\begin{figure}
\caption{\label{flux} Number of the desorption (top) and adsorption (bottom) events observed in the unit time sorted according to their time length. Only the data from the simulation of the heaviest polymers are shown. The curves follow approximately (multi-)lognormal relationships.}
\begin{center}
\includegraphics[width=8cm]{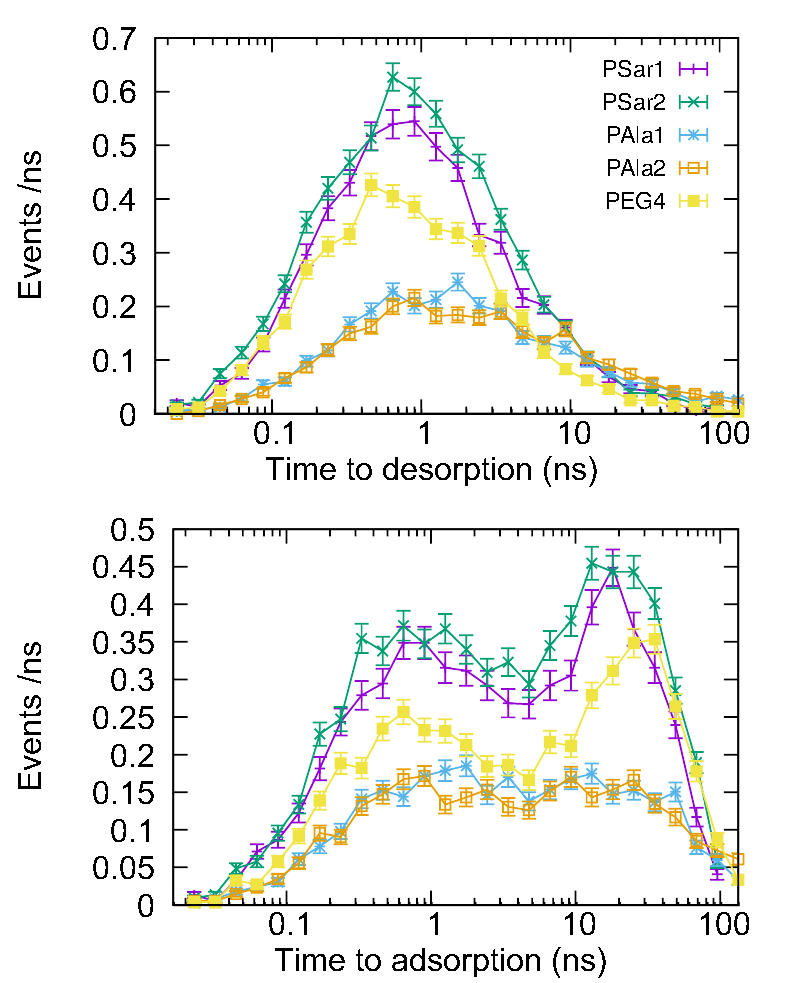}
\end{center}
\end{figure}

We, then, analyzed the kinetics of adsorption of the polymer molecules on the protein surface. An adsorption event was marked when more than half of the atoms of a desorbed molecule came closer than 0.5~nm to any protein heavy atom. On the contrary, a desorption event was defined when all the atoms of an adsorbed molecule reached farther than 0.5~nm from any protein heavy atom. The separation between the thresholds for the definition of adsorption and desorption events reduces the problem of recrossings \cite{Settanni2008, Radford2011, Buchete2008}. We then counted the number of adsorption and desorption events observed in the unit time sorted according to their duration (Tab.~\ref{ads_kin} and Fig.~\ref{flux}). The resulting distributions show approximately lognormal (or multi-lognormal) behavior with long tails for long event durations. Focusing on the simulations of the heaviest polymers (PEG$_7$, PSar and PAla), the data reveal that PSar simulations show the largest number of adsorption and desorption events in the unit time, which is expected given the higher concentration of polymer in these simulations. PEG$_7$ simulations show a proportionally smaller number of adsorption/desorption events compatible with the lower concentration. On the other hand, the low number of events registered for PAla which is lower than in PEG$_7$ simulations, is due to the lower diffusivity of PAla compared to the other polymers. Indeed, the diffusion coefficient of PAla derived from the mean square displacement in the simulations of pure polymer-water mixtures reached values below 31{\AA}$^2$/ns for the largest concentration tested, while in the case of PSar and PEG$_7$ the values remain above 42{\AA}$^2$/ns. The ultimate reason for the low diffusion coefficient of PAla and consequently the low number of adsorption/desorption events on the protein surface, is self aggregation. This point will be further discussed below. 

The distribution of the duration of the adsorption and desorption events (Fig.~\ref{flux}) also shows that polymer molecules remain adsorbed for relatively short times of the order of 1~ns before desorption, although for PAla the average is larger than in the other cases (Tab.~\ref{ads_kin}) and the tail at long times is thicker meaning that in that case several molecules can remain adsorbed for more than 100~ns. On the other hand the distribution of adsorption times has two peaks, the smallest one around the 1ns time scale, comparable with the one for desorption, and the largest one at the 10-30~ns time scale. The smallest peak is related to re-adsorption events, that is molecules that left the 0.5~nm layer around the protein only briefly before entering it again. The  second peak is related to polymer molecules reaching the protein surface after prolonged diffusion in the bulk region. The position of this last peak is related to the overall volume of the bulk region of the simulation box, which is approximately the same for the 5 simulations considered in Fig.~\ref{flux}.

\begin{table}
\caption{\label{ads_kin} Kinetics of adsorption}
\begin{threeparttable}
\begin{tabular}{lccccc}
\toprule
System & \specialcell{ Avg. num. \\ads. mol.} & \specialcell{ Avg. ads.\\ rate (ns$^{-1}$)} &\specialcell{ Avg. desorp.\\ rate (ns$^{-1}$)} & \specialcell{ Avg. Time \\to desorb. (ns)}\tnote{a} & \specialcell{ Avg. Time \\to adsorb. (ns)} \tnote{a}\\ 
\midrule
PEG1& 	32 $\pm$ 5 & 0.186 & 1.066 & 0.3 (+0.9 -0.2)&     1.0 (+7.0 -0.9)  	\\
PEG2& 	43 $\pm$ 6 & 0.181 & 1.074 & 0.3 (+0.9 -0.2)&     1.0 (+7.2 -0.9)  	\\
PEG3& 	42 $\pm$ 5 & 0.123 & 1.153 & 0.3 (+0.9 -0.2)&     1.2 (+10.1 -1.1) 	 \\
PEG4& 	13 $\pm$ 3 & 0.057 & 0.365 & 0.8 (+2.7 -0.6)&	   3.9 (+25.3 -3.4)	\\
PEG5& 	42 $\pm$ 6 & 0.091 & 1.257 & 0.3 (+0.9 -0.2)&     1.5 (+13.5 -1.3) 	\\ 
\midrule
PSar1& 	16 $\pm$ 4 & 0.062 & 0.285 & 1.2 (+3.9 -0.9)&	   4.7 (+23.7 -3.9)	\\ 
PSar2& 	18 $\pm$ 4 & 0.055 & 0.276 & 1.3 (+4.0 -1.0)&	   5.0 (+26.9 -4.2)	\\ 
\midrule
PAla1& 	23 $\pm$ 6 & 0.050 & 0.102 & 2.5 (+10.8 -2.0)&	   5.1 (+27.6 -4.3)	\\ 
PAla2& 	22 $\pm$ 7 & 0.039 & 0.103 & 2.7 (+12.1 -2.2)&	   5.5 (+33.9 -4.7)	\\
\bottomrule
\end{tabular}
\begin{tablenotes}
\item[a] The geometric mean and the 68\% confidence interval are reported.
\end{tablenotes}
\end{threeparttable}
\end{table}

We calculated the free energy of adsorption of the polymer molecules as $\Delta G_{ads}=-RT \log(\rho_{ads}\rho_{free})$ where $\rho_{ads}$ is the density of adsorbed polymer molecules, that is the number of adsorbed molecules divided by the volume of the 0.5~nm layer around the protein, and $\rho_{free}$ is the density of free molecules measured as the number of free molecules divided by the volume of the simulation box excluding protein and 0.5~nm adsorption layer. The data (Tab.~\ref{ads_ene}) confirm that PEG$_7$ and PSar have similar adsorption free energies for the protein surface while PAla adsorbs more strongly. In terms of enthalpic contributions in the present simulations (Tab.~\ref{ads_ene}) PAla shows overall larger attractive Van der Waals and electrostatic interactions for the protein surface than PSar because more molecules adsorb on the surface (Tab.~\ref{ads_kin}), even if the total concentration in the simulation box is lower. The interaction energies per adsorbed molecules (Tab.~\ref{ads_ene}), however, are similar for the three heaviest polymers studied here and are approximately equally distributed between electrostatic and Van der Waals interactions.

\begin{table}
\caption{\label{ads_ene} Energetics of adsorption}
\begin{threeparttable}
\begin{tabular}{lccccc}
\toprule
System & \specialcell{ $\Delta G_{ads}$\\  (Kcal/mol)} & \specialcell{Elec. total\\(kcal/mol)}& \specialcell{VdW total\\(kcal/mol)} & \specialcell{Elec. mol\\(kcal/mol)}& \specialcell{VdW mol\\(kcal/mol)} \\ 
\midrule
PEG1& 	-0.51 $\pm$ 0.10 & -376 $\pm$ 91  & -352 $\pm$ 59 & -11.6 $\pm$ 2.8 & -10.9 $\pm$ 1.8 	\\
PEG2& 	-0.48 $\pm$ 0.10 & -478 $\pm$ 101 & -449 $\pm$ 66 & -11.1 $\pm$ 2.4 & -10.4 $\pm$ 1.5	\\
PEG3& 	-0.39 $\pm$ 0.08 & -486 $\pm$ 98  & -427 $\pm$ 61 & -11.5 $\pm$ 2.3 & -10.1 $\pm$ 1.4 	 \\
PEG4& 	-0.66 $\pm$ 0.18 & -214 $\pm$ 82  & -215 $\pm$ 55 & -16.7 $\pm$ 6.4 & -16.8 $\pm$ 4.3 	\\
PEG5& 	-0.36 $\pm$ 0.09 & -461 $\pm$ 97  & -427 $\pm$ 65 & -11.0 $\pm$ 2.3 & -10.2 $\pm$ 1.6 	\\ 
\midrule
PSar1& 	-0.67 $\pm$ 0.17 & -260 $\pm$ 101 & -255 $\pm$ 59 & -16.5 $\pm$ 6.4 & -16.2 $\pm$ 3.8	\\ 
PSar2& 	-0.62 $\pm$ 0.15 & -303 $\pm$ 102 & -296 $\pm$ 60 & -16.5 $\pm$ 5.6 & -16.2 $\pm$ 3.2	\\ 
\midrule
PAla1& 	-1.15 $\pm$ 0.23 & -370 $\pm$ 118 & -369 $\pm$ 109 & -16.2 $\pm$ 5.2 & -16.2 $\pm$ 4.8   \\ 
PAla2& 	-1.00 $\pm$ 0.27 & -380 $\pm$ 155 & -347 $\pm$ 115 & -17.1 $\pm$ 7.0 & -15.7 $\pm$ 5.2   \\
\bottomrule
\end{tabular}
\end{threeparttable}
\end{table}

\begin{figure}
\caption{\label{rdf_pala_clu} A) Cumulative radial distribution function of water oxygen atoms around the heavy atoms of the polymers (data from 3 selected runs). B) Cartoon of a PAla cluster formed in the simulation of the polymer-water mixture water molecules are omitted for clarity). The black connectors indicate hydrogen bonds.}
\begin{center}
\includegraphics[width=8cm]{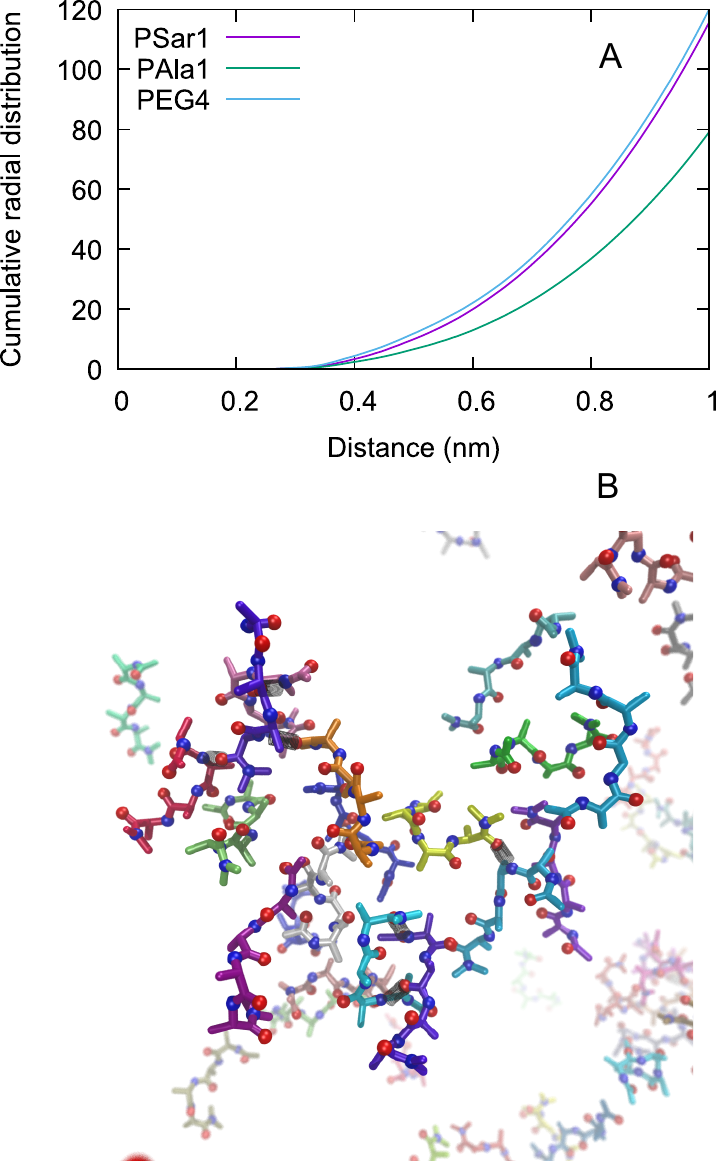}
\end{center}
\end{figure}

All the systems that we have studied include basically three interacting partners, the protein, the polymer and water (here for simplicity we exclude the ions). The data presented so far show that the interaction pattern between protein surface and polymers are similar in the three cases both in terms of the amino acid types involved in the interactions and in terms of the average interaction energy of the single polymer molecules with the surface. The data also show that the shape of the polymers is only marginally affected by the adsorption on the protein surface (only PAla showed a mild shift to more compact structures). We, then  considered the interactions between the polymers and water. From the simulations we have measured the amount of hydrogen bonds made by each polymer with the surrounding water and have concluded that PEG$_7$, PSar and PAla have a similar accessible surface area (between 560 and 567$\pm$28 {\AA}$^2$) and make a similar number of h-bonds per molecule with water (between 2.66 and 2.84$\pm$0.18). The similarity in the hydrophilicity between PEG and PSar is in agreement with experimental findings on the elution properties and contact angle measurements of PSar~\cite{Lau2012}. On the other hand, the cumulative radial distribution function of water around the heavy atoms of the simulated polymers (Fig~\ref{rdf_pala_clu}A) shows that Psar atoms have in between 1.2 and 1.5 times more water molecules in a 1nm shell around them than Pala atoms. The radial distribution function is calculated for each heavy atom of the polymer and averaged over all the heavy atoms of the molecule. As such, it is sensitive to the size and shape of the molecule. So we limited our comparison to PAla, PSar and PEG$_7$  which have 26 , 26 and 22 atoms, respectively.  This means that each atom in PEG$_7$  "sees" slightly less polymer atoms in the 1nm shell than PSar atoms and consequently slightly more water molecules, as shown in Fig~\ref{rdf_pala_clu}A. On the other hand, PAla and PSar have the same number of heavy atoms and although PAla is slightly more compact on average than PSar (see above and Fig~\ref{shape_comp}) this does not justify the dramatic drop of water content in the 1nm shell around the atoms of PAla with respect to PSar. The latter is due to the fact that PAla tends to form large clusters which contain few water molecules. The clusters (containing sometimes more than ten PAla molecules) are stabilized by a combination of inter-molecular beta-bridges (pairs of hydrogen-bonded peptides in beta conformation) and hydrophobic interactions between the side-chains (Fig.~\ref{rdf_pala_clu}B). These structures do not form in PSar where inter-molecular beta-bridges are prevented both by the lack of strong hydrogen bond donors on the backbone and by the larger conformational flexibility provided by a Ramachandran angle distribution~\cite{Ramachandran1963} more similar to glycine than to alanine~\cite{sarcosineff}, confirmed recently by cluster analysis of simulations of single polymers in solution~\cite{Prakash2018}. 

\section{Conclusions}
Here, we have used MD simulations to assess how two hydrophilic polymers of biotechnological importance, because of their possible use as coatings of therapeutic nanoparticles, interact with the surface of HSA, an important blood protein. The simulations help and explain that PEG and PSar develop a very similar interaction pattern with the protein surface, both in terms of the affinity with the various amino acids and in terms of overall intensity of the interaction. Adsorption on the protein surface is a reversible process which occurs at a relatively fast rate. The data show also that the structure of these polymers is not modified substantially during the adsorption process. As a consequence of this, the thickness of the adsorption layer is slightly larger for PSar than PEG, reflecting its bulkier and less flexible structure. In comparison, PAla, a polymer isomer of PSar known for its propensity to self-aggregate and induce protein aggregation, shows a substantially higher affinity for the protein surface, which is not due to a larger interaction energy with the protein or a different pattern of interaction with the surface amino acids but rather due to the way the polymer interacts with water, and, in particular, due to the tendency of the polymer to reduce the surface exposed to water either by self-aggregating or by adsorbing to the protein surface.

\section*{Acknowledgements}
TS gratefully acknowledges financial support from the Graduate School Materials Science in Mainz.
GS gratefully acknowledges financial support from the Max-Planck Graduate Center with the University of Mainz.
We gratefully acknowledge support with computing time from the HPC facility Hazelhen at the High performance computing center Stuttgart (project Flexadfg) and the HPC facility Mogon at the university of Mainz. This work was supported by the German Science Foundation within SFB 1066 project Q1.



\section*{References}





\end{document}